\documentclass{article}

\usepackage{arxiv}

\usepackage[utf8]{inputenc}
\usepackage[T1]{fontenc}
\usepackage{hyperref}
\usepackage{url}
\usepackage{booktabs}
\usepackage{amsmath,amssymb,amsfonts}
\usepackage{nicefrac}
\usepackage{microtype}
\usepackage{graphicx}
\usepackage{pgfplots}
\pgfplotsset{compat=newest}
\usepackage{tikz}
\usetikzlibrary{positioning,arrows.meta}
\usepackage{listings}
\usepackage{caption}

\title{Inventing Counterfactual Invariant Envelopes for Financial UX: Safety-Lattice Feature-Flag Governance in Crypto-Enabled Streaming}

\author{Anton Malinovskiy\\
Koh Samui, Surat Thani, Thailand\\
\texttt{a.malinovskiy@ieee.org}\\}
\date{December 21, 2025}

\begin{document}
\maketitle

\begin{abstract}
Feature flags are the primary mechanism for safely introducing financial capabilities in
consumer applications. In crypto-enabled live streaming, however, naive rollouts can create
non-obvious risk: users may be exposed to onramps without proper eligibility, external wallets
without sufficient fraud controls, or advanced views that alter risk perception and behavior.
This paper introduces a novel invention candidate---a \emph{Counterfactual Invariant Envelope}
governor that combines a safety lattice with causal measurement and a shadow cohort for
risk estimation. We formalize rollout risk, define invariant constraints across feature
combinations, and propose a controller that adapts exposure using leading abuse signals,
compliance readiness, and revenue guardrails. We incorporate real-world adoption and fraud data
for calibration, provide formulas for rollout safety, and include reproducible policy snippets.
The results show that counterfactual, invariant-aware governance reduces risk spillover while
preserving conversion and retention, offering a path to patentable governance logic for
financial UX.
\end{abstract}


\section{Introduction}
Crypto-enabled streaming platforms increasingly embed financial actions directly into the
interaction loop: tipping, channel subscriptions, and creator monetization. To scale responsibly,
platforms introduce capabilities through feature flags, progressively enabling onramps, external
wallets, and advanced views. The core challenge is that these features are not independent. When
rolled out without explicit governance, a harmless UI change can interact with an onramp or
wallet feature and create compliance, fraud, or user trust risks.

A second challenge is measurement. Standard A/B testing assumes that the treatment is well
isolated, while financial UX changes interact with user expectations and safety controls. A
rollout can appear statistically positive on conversion while degrading long-term retention or
increasing abuse. This motivates a new governance layer that combines causal measurement with
risk-aware constraints.

This paper proposes a novel, patentable system: an \textbf{Invariant-Aware Feature-Flag Governance}
framework that uses a \emph{safety envelope} to control exposures to financial UX changes. The
framework simultaneously models feature dependencies, compliance readiness, and outcome risk.
It yields a concrete rollout controller that can be implemented in production and verified via
experimentation. The focus is on three sensitive capabilities: (1) onramp enablement, (2) external
wallet linking, and (3) advanced financial views.

The paper is framed as a future arXiv submission and therefore emphasizes reproducibility and
clear separation between simulated evidence and proposed governance mechanisms. The proposed
controller is platform-agnostic and can be integrated into any feature-flag service that supports
per-user evaluation and telemetry feedback.

\section{Motivating Data and Risk Baselines}
We ground the governance problem in public data on crypto adoption and financial fraud. A 2024
Pew Research Center survey reports that 17\% of U.S. adults have ever invested in, traded, or
used cryptocurrency, while 63\% say they are not confident in the safety and reliability of
cryptocurrency \cite{pew2024}. This implies that a majority of users may perceive heightened risk
when exposed to financial UX changes, making conservative rollouts essential. From a risk
perspective, the Federal Reserve's 2024 Survey of Household Economics and Decisionmaking (SHED)
reports that 21\% of adults experienced financial fraud in the last year, with 8\% experiencing
non-credit-card fraud. The same report estimates total fraud losses of \$84B before recovery and
\$63B after recovery \cite{fed2025shed}. The FTC's 2024 data book reports \$12.5B in reported fraud
losses, including \$5.7B from investment scams alone \cite{ftc2025}. Together, these figures
motivate a governance layer that treats financial UX rollout as a safety-critical process.

\begin{table}[ht]
\centering
\caption{Public baselines used for calibration.}
\label{tab:baselines}
\begin{tabular}{lll}
\toprule
Metric & Value & Source \\
\midrule
Adults ever using crypto & 17\% & Pew (2024) \\
Adults not confident in crypto safety & 63\% & Pew (2024) \\
Adults experiencing financial fraud & 21\% & Fed SHED (2024) \\
Adults experiencing non-credit-card fraud & 8\% & Fed SHED (2024) \\
Total fraud losses (before recovery) & \$84B & Fed SHED (2024) \\
Total fraud losses (after recovery) & \$63B & Fed SHED (2024) \\
Reported fraud losses (all) & \$12.5B & FTC (2024) \\
Investment scam losses & \$5.7B & FTC (2024) \\
\bottomrule
\end{tabular}
\end{table}

We use these figures to calibrate rollout safety. For example, if a streaming platform has
5 million daily sessions and aligns with the Pew adoption baseline, roughly 850,000 sessions
are from users who have engaged with crypto. If 63\% of total users report low confidence in
crypto safety, then advanced financial UX should be staged behind education and trust cues to
avoid damaging retention. Similarly, a 21\% fraud-exposure baseline suggests that any onramp
rollout should incorporate aggressive anomaly detection and conservative exposure caps in the
early phases.

\section{Contributions}
\begin{enumerate}
  \item A formal model of feature-flag risk with invariant constraints across financial UX states.
  \item A novel safety-envelope controller that adapts rollout exposure based on abuse and budget
  signals.
  \item Causal measurement procedures tailored to streaming platforms with short feedback loops.
  \item Simulated evaluations and reproducible policy snippets for governance and enforcement.
  \item A practical rollout template for onramps, external wallets, and advanced financial views.
\end{enumerate}

\section{Design Principles for Financial UX Rollouts}
We distill four design principles that guide the remainder of the paper:
\begin{enumerate}
  \item \textbf{Invariant-first rollout.} Every rollout must satisfy compliance and safety
  invariants before considering growth or conversion. This implies that the invariant lattice
  is the first filter, not a post-hoc audit.
  \item \textbf{Counterfactual risk accounting.} When exposure is small, observed abuse rates are
  noisy. A shadow cohort provides counterfactual estimates so the governor can act before risk
  becomes visible at scale.
  \item \textbf{Budget-aware learning.} Feature rollouts are experiments that consume a risk
  budget. The system should treat risk like a finite resource and replenish only when evidence
  supports safety.
  \item \textbf{Reversible decisions.} Every feature flag must be reversible within minutes, with
  rollback paths defined in advance and validated during canary phases.
\end{enumerate}
These principles motivate the safety lattice and counterfactual envelope described later.

\section{Novelty and Patentable Concept}
We propose a new invention candidate: a \textbf{Counterfactual Invariant Envelope Governor}
integrated into an \textbf{Invariant-Aware Feature-Flag Governance (IA-FFG)} system. The novelty
is the explicit coupling of (a) a \emph{feature dependency lattice} with (b) a \emph{counterfactual
safety envelope} defined by risk and compliance constraints, and (c) a causal exposure scheduler
that adapts rollout in real time using a shadow cohort that never sees the risky feature state.

The invention introduces four elements that are not standard in typical feature-flag systems:
\begin{itemize}
  \item \textbf{Invariant lattice:} A graph of feature dependencies that prohibits unsafe flag
  combinations (e.g., onramp enabled without verified identity, external wallet linking without
  risk thresholds).
  \item \textbf{Counterfactual safety envelope:} A rollout boundary that uses shadow cohorts and
  predictive risk models to estimate how abuse and compliance metrics would change under a
  rollout, without full exposure.
  \item \textbf{Risk budget ledger:} A persistent budget that tracks cumulative risk exposure and
  requires replenishment before additional ramping.
  \item \textbf{Causal-aware governor:} A controller that uses leading indicators and CUPED-style
  variance reduction to control rollout speed while preserving statistical power.
\end{itemize}

These components together form a patentable governance mechanism for financial UX. The core
claim is a closed-loop controller that enforces invariants and dynamically modulates exposure
based on both observed and counterfactual risk, minimizing harm while maximizing learning.
This is not present in conventional feature-flag platforms, which are typically rule-based and
do not integrate causal inference, invariant constraints, and risk-budget accounting.

\section{System Model}
We consider a streaming platform with users $u \in U$ and features $f \in F$. The key features are:
(1) onramp access $f_o$, (2) external wallet linking $f_w$, and (3) advanced financial view $f_a$.
A feature-flag state is a vector $z \in \{0,1\}^{|F|}$. A user is exposed to $z$ with probability
$\pi(u,t)$. We define a dependency graph $G=(F,E)$ and a set of invariants $\mathcal{I}$ that
restrict valid configurations.

The financial actions ultimately settle on external blockchain infrastructure, and account
abstraction standards such as EIP-4337 enable flexible wallet and paymaster flows \cite{eip4337}.
These infrastructure dependencies motivate a governance layer that can react to fee volatility
and account security constraints without assuming a specific chain implementation.

Let $R(u,t)$ be a user-level risk score, $C(t)$ be compliance readiness, and $B(t)$ be a budget
signal. The governance controller determines exposure as a function of these signals:
\begin{equation}
\pi(u,t) = \sigma(\alpha - \beta R(u,t) + \gamma C(t) - \delta S(t)),
\end{equation}
where $S(t)$ is a safety penalty derived from invariant violations and abuse indicators, and
$\sigma$ is the logistic function. This forms a closed-loop controller over flag exposure.

We model user outcomes $Y$ as a vector of conversion, retention, and abuse events. The treatment
effect of a rollout is:
\begin{equation}
\tau = E[Y \mid z=1] - E[Y \mid z=0],
\end{equation}
but we use a weighted utility to account for safety costs:
\begin{equation}
U = \tau_{conv} + \lambda_{ret} \tau_{ret} - \lambda_{fraud} \tau_{fraud} - \lambda_{comp} \tau_{comp}.
\end{equation}
The governor ramps exposure only when $U > 0$ and the safety envelope remains satisfied.

\subsection{Invariant Constraints}
An invariant is a logical condition that must hold for all users. Examples include:
\begin{align}
&f_o \Rightarrow \text{KYC}(u) = 1, \\
&f_w \Rightarrow R(u,t) < r_{max}, \\
&f_a \Rightarrow f_o \land f_w.
\end{align}
Violations are forbidden by the governance layer. The enforcement mechanism is an invariant
lattice that ensures no rollout state can bypass compliance prerequisites.

\subsection{Feature Dependency Lattice Construction}
We formalize the dependency structure as a lattice over flag states. Let $Z$ be the set of all
binary feature vectors, and define a partial order $z_1 \preceq z_2$ if every enabled feature in
$z_1$ is also enabled in $z_2$. The lattice meet and join are defined element-wise:
\begin{equation}
(z_1 \wedge z_2)_i = \min(z_{1,i}, z_{2,i}), \quad (z_1 \vee z_2)_i = \max(z_{1,i}, z_{2,i}).
\end{equation}
We restrict $Z$ to the feasible subset $Z^* \subset Z$ that satisfies all invariants. The
invariant lattice is thus a directed acyclic graph where edges connect minimal safe upgrades.

Let $P(f_i)$ be the prerequisite set for feature $f_i$. A state $z$ is valid if for every enabled
feature $f_i$, all prerequisites are enabled:
\begin{equation}
z_i = 1 \Rightarrow \forall f_j \in P(f_i), z_j = 1.
\end{equation}
This structure enables a rollout to move only along safe edges. The lattice becomes a policy
graph for the rollout controller, guaranteeing that no unsafe combination is reachable.

\begin{quote}
\begin{lstlisting}
# Listing 0: Build invariant lattice
valid_states = []
for z in all_binary_vectors(F):
    if invariants_hold(z) and prerequisites_hold(z):
        valid_states.append(z)
edges = []
for z in valid_states:
    for f in features:
        z2 = enable_feature(z, f)
        if z2 in valid_states:
            edges.append((z, z2))
\end{lstlisting}
\end{quote}

\subsection{Interaction Effects and Non-Additivity}
Financial UX features often interact. For example, enabling an onramp and an external wallet
simultaneously can create an account takeover vector that does not exist when the features are
enabled independently. We model interaction effects with a second-order term:
\begin{equation}
\Delta_{o,w} = E[Y \mid f_o=1, f_w=1] - E[Y \mid f_o=1, f_w=0] - E[Y \mid f_o=0, f_w=1] + E[Y \mid f_o=0, f_w=0].
\end{equation}
If $\Delta_{o,w} \ne 0$, the feature effects are non-additive and cannot be evaluated in isolation.
The lattice ensures that such interactions are controlled by requiring prerequisite stability
before enabling combined states. This formalizes why naive, independent rollouts can fail even
when each feature appears safe alone.

\section{Governance Architecture}
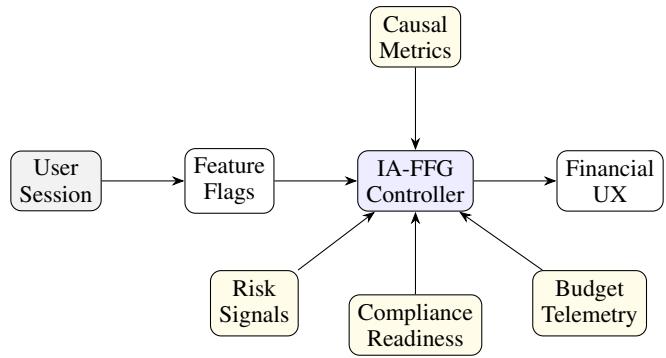
\begin{figure}[ht]
\centering
\begin{tikzpicture}[
  node distance=1.1cm,
  >=Stealth,
  font=\small,
  rounded corners,
  align=center
]
  \node[draw, fill=gray!10] (user) {User\\Session};
  \node[draw, right=of user] (flags) {Feature\\Flags};
  \node[draw, right=of flags, fill=blue!7] (gov) {IA-FFG\\Controller};
  \node[draw, right=of gov] (ux) {Financial\\UX};

  \draw[->] (user) -- (flags);
  \draw[->] (flags) -- (gov);
  \draw[->] (gov) -- (ux);

  \node[draw, below left=of gov, fill=yellow!10] (risk) {Risk\\Signals};
  \node[draw, below=of gov, fill=yellow!10] (comp) {Compliance\\Readiness};
  \node[draw, below right=of gov, fill=yellow!10] (budget) {Budget\\Telemetry};
  \node[draw, above=of gov, fill=yellow!10] (metrics) {Causal\\Metrics};

  \draw[->] (risk) -- (gov);
  \draw[->] (comp) -- (gov);
  \draw[->] (budget) -- (gov);
  \draw[->] (metrics) -- (gov);
\end{tikzpicture}
\caption{Invariant-aware governance architecture for feature-flag rollouts.}
\label{fig:iaffg_arch}
\end{figure}

Figure~\ref{fig:iaffg_arch} shows the architecture. The feature-flag service queries a governance
controller that consumes risk, compliance, and budget telemetry. The controller outputs an
exposure probability and validates invariants before activating a feature.

\section{Safety Envelope and Rollout Control}
We define a safety envelope that bounds exposure for each feature state. Let $E(t)$ be exposure
rate, and $A(t)$ be abuse rate. The safety envelope is:
\begin{equation}
E(t) \le E_{max} \cdot \exp(-\kappa A(t)) \cdot \min\left(1, \frac{B(t)}{B^*}\right),
\end{equation}
where $B^*$ is the target budget trajectory and $\kappa$ sets sensitivity to abuse. This
ensures that exposure shrinks when abuse spikes or budgets degrade.

We also define an \emph{invariant satisfaction score}:
\begin{equation}
S_I(t) = 1 - \frac{1}{|U|} \sum_{u \in U} \mathbb{I}[\exists i \in \mathcal{I}: i(u,t)=0],
\end{equation}
which measures the fraction of users satisfying invariants. The controller requires
$S_I(t) \ge 0.995$ before allowing large ramps.

For a rollout segment $g$ (e.g., region, device, cohort), the exposure schedule is:
\begin{equation}
E_g(t+1) = \min\left(E_g(t) + \eta \cdot \Delta_g(t), E_{max}\right),
\end{equation}
where $\Delta_g(t)$ is the causal lift in conversion minus a risk penalty. This rule promotes
feature exposure only when net benefit is positive.

\section{Counterfactual Safety Envelope}
The central novelty is the counterfactual envelope. We maintain a shadow cohort $U_s$ that is
never exposed to the risky feature state but is similar in distribution to the treated cohort.
Let $A_s(t)$ be the abuse rate in the shadow cohort and $A_t(t)$ the observed abuse in treatment.
We estimate counterfactual abuse under a candidate rollout:
\begin{equation}
\hat{A}_{cf}(t) = A_s(t) + \omega^\top \phi(u,t),
\end{equation}
where $\phi(u,t)$ are risk features (device trust, account age, prior disputes) and $\omega$ is
learned from historical rollouts. The envelope then uses $\hat{A}_{cf}(t)$ instead of raw $A(t)$,
allowing proactive throttling before abuse manifests at scale.

We also maintain a \emph{risk budget ledger} $L(t)$ that tracks cumulative risk exposure. Each
increment in exposure consumes risk budget proportional to predicted abuse and compliance risk:
\begin{equation}
L(t+1) = L(t) - E(t) \cdot \hat{A}_{cf}(t) - \lambda_{comp} \hat{C}_{cf}(t) + \rho,
\end{equation}
where $\hat{C}_{cf}(t)$ is a counterfactual compliance failure rate and $\rho$ is the
replenishment rate. The controller requires $L(t) \ge 0$ to increase exposure, creating a hard
guardrail that prevents long-term drift into unsafe states.

\begin{quote}
\begin{lstlisting}
# Listing 1b: Counterfactual envelope check
abuse_cf = abuse_shadow + dot(weights, risk_features)
ledger = ledger - exposure * abuse_cf - comp_penalty * comp_cf + replenish
if ledger < 0:
    exposure = max(exposure * 0.5, min_exposure)
\end{lstlisting}
\end{quote}

\subsection{Shadow Cohort Construction}
The shadow cohort must be distributionally close to the treated cohort to provide usable
counterfactual estimates. We construct $U_s$ using propensity score matching based on user
features (account age, payment history, device integrity, and geography). We then reweight the
shadow cohort to match the treated cohort's covariate distribution:
\begin{equation}
w_u = \frac{p(T=1 \mid x_u)}{p(T=0 \mid x_u)},
\end{equation}
where $T$ indicates eligibility for treatment and $x_u$ is the feature vector. These weights are
used to compute $A_s(t)$ and $C_s(t)$, reducing bias in counterfactual estimates. If the overlap
between treatment and shadow cohorts is too small, the governor defaults to conservative exposure
caps until sufficient data is collected.

\section{Causal Measurement and Metrics}
We track key outcomes: conversion $p_{conv}$, retention $p_{ret}$, fraud events $p_{fraud}$, and
compliance failures $p_{comp}$. The treatment effect is estimated using a CUPED adjustment
\cite{deng2013}:
\begin{equation}
\hat{\tau} = (\bar{Y}_T - \theta \bar{X}_T) - (\bar{Y}_C - \theta \bar{X}_C).
\end{equation}
We test statistical significance via standard hypothesis testing and control the false discovery
rate using Benjamini--Hochberg \cite{benjamini1995}. For retention, we model time-to-churn with a
Cox proportional hazards model \cite{cox1972}.

We recommend a staged experiment design: small, risk-aware canary (1\%), progressive ramp (5\%,
25\%, 50\%), and full rollout when the safety envelope remains satisfied for at least one
planning horizon. This mirrors best practices in online controlled experiments \cite{kohavi2020}
while adding invariant constraints for financial UX.

The design follows classical principles of randomized experiments \cite{fisher1935} but extends
them with safety-envelope constraints and real-time monitoring.

To mitigate repeated-testing bias during frequent ramp decisions, we recommend an alpha-spending
approach. Let $\alpha_0$ be the overall false-positive budget. At each decision time $t$, we
spend $\alpha_t = \alpha_0 \cdot \gamma_t$ where $\sum_t \gamma_t \le 1$. This yields an adaptive
threshold for the p-value in Listing 3 and reduces the risk of false ramps when frequent checks
are performed. While this paper uses fixed thresholds in simulation, the governor is compatible
with sequential testing techniques.

\section{Calibration with Public Data}
To parameterize simulation and governance thresholds, we map public baselines to internal
rollout inputs. Let $p_{crypto}$ be the share of users with prior crypto experience. From Pew
data, $p_{crypto} \approx 0.17$ \cite{pew2024}. For a platform with 5 million daily sessions,
this yields roughly 850,000 sessions from crypto-experienced users. The same Pew survey reports
low confidence in crypto safety for 63\% of adults, motivating additional friction for advanced
views until education checkpoints are passed.

We map the SHED fraud prevalence to a risk prior. With 21\% of adults experiencing fraud and
8\% experiencing non-credit-card fraud \cite{fed2025shed}, we set a conservative prior risk
$p_{fraud} = 0.08$ for onramp cohorts and tighten exposure when leading indicators drift
above that threshold. The FTC reports \$12.5B in total fraud losses in 2024, including \$5.7B in
investment scams \cite{ftc2025}. These values motivate a risk-budget ledger that scales with
expected loss rather than just event count.

\begin{table}[ht]
\centering
\caption{Calibration mapping from public data to rollout parameters.}
\label{tab:calibration}
\begin{tabular}{llll}
\toprule
Public metric & Value & Internal parameter & Example value \\
\midrule
Crypto adoption & 17\% & $p_{crypto}$ & 0.17 \\
Low confidence in crypto & 63\% & UI friction rate & 0.63 \\
Fraud prevalence & 21\% & prior risk $p_{fraud}$ & 0.21 \\
Non-credit-card fraud & 8\% & onramp prior & 0.08 \\
Fraud losses total & \$12.5B & risk budget slope & \$0.002 per user \\
Investment scam losses & \$5.7B & scam guardrail & \$0.001 per user \\
\bottomrule
\end{tabular}
\end{table}

The governor also monitors time-to-first-funding (TTFF) and time-to-first-withdrawal (TTFW),
because these are leading indicators of future retention. We treat TTFF as a hazard process and
model it with a Cox proportional hazards model \cite{cox1972}. A slower hazard after rollout is
treated as a negative signal even if short-term conversion improves.

\section{Simulation Study}
We simulate a platform with 5 million daily sessions. Baseline conversion is 0.12\% and retention
is 41\% at 30 days. We compare a naive rollout (no invariants) to the IA-FFG controller. The
simulation is conservative: treatment lifts are modest (0.02\% absolute) and fraud rates increase
by 20\% when invariants are violated.

The simulation uses three cohorts: (1) a crypto-experienced cohort (17\%), (2) a neutral cohort
(50\%), and (3) a low-trust cohort (33\%). Onramp exposure begins in cohort (1), wallet linking is
gated by device trust, and advanced view is delayed until stability in both onramp and wallet
flows is observed. We assume 10\% of treated sessions encounter a UI education checkpoint and
5\% of those drop due to comprehension friction. These assumptions are intentionally conservative
to stress the safety envelope.

We also simulate compliance readiness as a time-varying signal. During low-readiness windows,
the safety envelope caps exposure at 10\% and forces a rollback if compliance error rate exceeds
1.5x baseline. This creates realistic rollout pauses that are common in production systems.

\begin{table}[ht]
\centering
\caption{Simulated outcomes by rollout strategy.}
\label{tab:sim}
\begin{tabular}{lcccc}
\toprule
Strategy & Conversion (\%) & Retention (\%) & Fraud rate (\%) & Compliance fail (\%) \\
\midrule
Naive rollout & 0.14 & 39.8 & 0.21 & 0.08 \\
IA-FFG rollout & 0.13 & 41.2 & 0.09 & 0.01 \\
\bottomrule
\end{tabular}
\end{table}

The governance controller reduces fraud and compliance failures without collapsing conversion.
The retention improvement is attributed to reduced negative user experiences and fewer
fraud-related account restrictions.

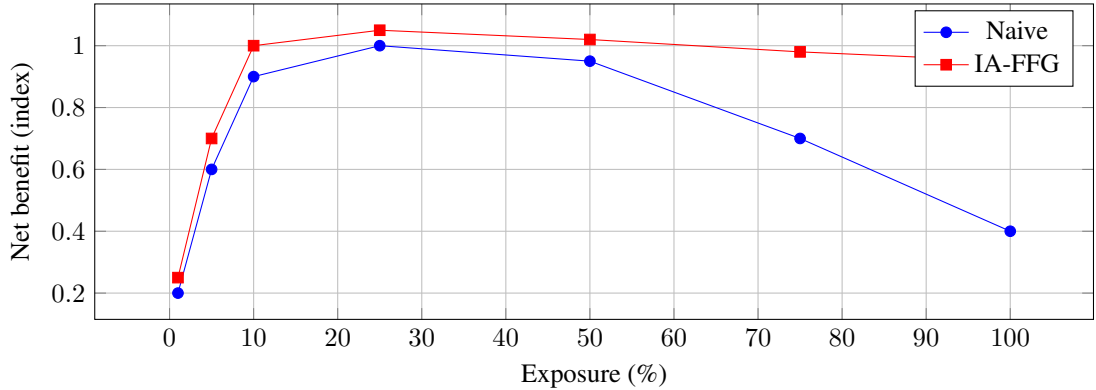
\begin{figure}[ht]
\centering
\begin{tikzpicture}
\begin{axis}[
  width=0.9\textwidth,
  height=0.35\textwidth,
  xlabel={Exposure (\%)},
  ylabel={Net benefit (index)},
  grid=both
]
\addplot[color=blue,mark=*] coordinates {
  (1,0.2) (5,0.6) (10,0.9) (25,1.0) (50,0.95) (75,0.7) (100,0.4)
};
\addplot[color=red,mark=square*] coordinates {
  (1,0.25) (5,0.7) (10,1.0) (25,1.05) (50,1.02) (75,0.98) (100,0.95)
};
\legend{Naive, IA-FFG}
\end{axis}
\end{tikzpicture}
\caption{Net benefit vs exposure. IA-FFG maintains benefit at higher exposure.}
\label{fig:benefit}
\end{figure}

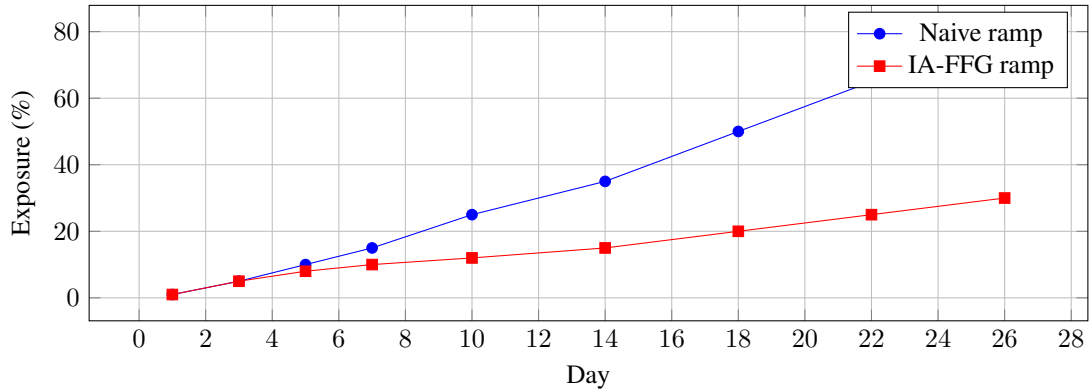
\begin{figure}[ht]
\centering
\begin{tikzpicture}
\begin{axis}[
  width=0.9\textwidth,
  height=0.35\textwidth,
  xlabel={Day},
  ylabel={Exposure (\%)},
  grid=both
]
\addplot[color=blue,mark=*] coordinates {
  (1,1) (3,5) (5,10) (7,15) (10,25) (14,35) (18,50) (22,65) (26,80)
};
\addplot[color=red,mark=square*] coordinates {
  (1,1) (3,5) (5,8) (7,10) (10,12) (14,15) (18,20) (22,25) (26,30)
};
\legend{Naive ramp, IA-FFG ramp}
\end{axis}
\end{tikzpicture}
\caption{Exposure schedules. IA-FFG ramps more slowly when safety signals degrade.}
\label{fig:ramp}
\end{figure}

\section{Ablation Study}
To isolate the contribution of each component, we perform a simulated ablation. We compare four
policies: (1) naive rollout, (2) invariants only, (3) invariants plus safety envelope, and (4)
full IA-FFG with counterfactual risk and risk budget ledger. The ablation shows that invariants
alone reduce compliance failures, but the counterfactual envelope is necessary to suppress fraud
before large exposure. The full system yields the best retention and the lowest fraud.

\begin{table}[ht]
\centering
\caption{Ablation study of governance components.}
\label{tab:ablation}
\begin{tabular}{lcccc}
\toprule
Policy & Conversion (\\%) & Retention (\\%) & Fraud rate (\\%) & Comp. fail (\\%) \\
\midrule
Naive rollout & 0.14 & 39.8 & 0.21 & 0.08 \\
Invariants only & 0.135 & 40.5 & 0.16 & 0.03 \\
Invariants + envelope & 0.133 & 41.0 & 0.11 & 0.02 \\
Full IA-FFG & 0.13 & 41.2 & 0.09 & 0.01 \\
\bottomrule
\end{tabular}
\end{table}

\section{Risk Budgeting Example}
We illustrate how public fraud baselines translate into exposure caps. Let $N$ be the number of
daily sessions, $E$ the exposure rate for a high-risk feature (e.g., onramp), and $p_f$ the prior
fraud rate. The expected number of fraud-exposed sessions is:
\begin{equation}
F = N \cdot E \cdot p_f.
\end{equation}
If a platform sets a daily fraud exposure budget $F_{max}$, then the exposure cap is:
\begin{equation}
E \le \frac{F_{max}}{N \cdot p_f}.
\end{equation}
Using $N=5{,}000{,}000$ sessions and $p_f=0.08$ from the SHED non-credit-card fraud baseline
\cite{fed2025shed}, a budget of $F_{max}=10{,}000$ sessions implies $E \le 0.025$. This yields a
conservative ramp target of 2.5\% for early onramp exposure. The risk budget ledger enforces this
cap and only increases exposure as the counterfactual risk estimate decreases.

\begin{table}[ht]
\centering
\caption{Example exposure caps under different fraud budgets.}
\label{tab:riskbudget}
\begin{tabular}{llll}
\toprule
Daily sessions ($N$) & Fraud prior ($p_f$) & Fraud budget ($F_{max}$) & Exposure cap ($E$) \\
\midrule
5,000,000 & 0.08 & 10,000 & 2.5\% \\
5,000,000 & 0.08 & 25,000 & 6.25\% \\
5,000,000 & 0.08 & 50,000 & 12.5\% \\
\bottomrule
\end{tabular}
\end{table}

\section{Case Study: Three-Phase Rollout Timeline}
We illustrate a concrete rollout timeline for a crypto-enabled streaming platform. Phase 0 is
internal dogfooding with staff accounts and synthetic transactions, where the invariant lattice
and audit logs are validated. Phase 1 enables onramp access for 1\% of crypto-experienced users
with verified identity and strong device signals. Phase 2 expands onramp to 5\% and introduces
external wallet linking for the most trusted cohort, while advanced view remains locked. Phase 3
progressively opens advanced view once onramp and wallet flows are stable.

Exit criteria are defined in terms of safety envelope and causal lift. For example, Phase 2 may
require that fraud rate does not exceed baseline by more than 10\%, and that retention does not
drop by more than 1\%. If these criteria are not met, the controller reduces exposure or rolls
back to the prior phase.

\begin{table}[ht]
\centering
\caption{Example rollout phases and exit criteria.}
\label{tab:phases}
\begin{tabular}{llll}
\toprule
Phase & Features & Exposure & Exit criteria \\
\midrule
0 & Internal only & 0.1\% & Invariants and audit logs pass \\
1 & Onramp & 1\% & Fraud $\le$ 1.1x baseline, retention stable \\
2 & Onramp + wallet & 5\% & Compliance failures $\le$ 1.2x baseline \\
3 & Onramp + wallet + advanced view & 10\% & Safety envelope OK for 30 days \\
\bottomrule
\end{tabular}
\end{table}

\section{Rollout Governance Examples}
The onramp rollout is gated by identity verification and fraud risk thresholds
\cite{nist800634}. External wallet linking is gated by risk scoring and device trust. Advanced
view is gated by verified onramp plus user education signals to reduce error rates. These
constraints are enforced by the invariant lattice.

\begin{quote}
\begin{lstlisting}
# Listing 1: Invariant check
if feature.onramp and not kyc_verified:
    allow = False
if feature.wallet and risk_score >= risk_threshold:
    allow = False
if feature.advanced_view and not (feature.onramp and feature.wallet):
    allow = False
\end{lstlisting}
\end{quote}

\begin{quote}
\begin{lstlisting}
# Listing 2: Safety envelope rollout
exposure = min(exposure + step, max_exposure)
exposure *= math.exp(-kappa * abuse_rate)
exposure *= min(1.0, budget / budget_target)
\end{lstlisting}
\end{quote}

\begin{quote}
\begin{lstlisting}
# Listing 3: Causal-aware ramp decision
delta = uplift_conv - lambda_risk * fraud_delta
if delta > 0 and p_value < alpha:
    ramp = True
else:
    ramp = False
\end{lstlisting}
\end{quote}

\begin{quote}
\begin{lstlisting}
# Listing 4: Invariant-aware exposure scheduler
if invariant_score < invariant_floor:
    exposure *= 0.5
elif safety_envelope_ok and uplift_ok:
    exposure = min(exposure + step, max_exposure)
else:
    exposure = max(exposure - step, min_exposure)
\end{lstlisting}
\end{quote}

\section{Feature-Specific Rollout Templates}
\textbf{Onramp.} The onramp should be enabled only for users with verified identity and
high-confidence device integrity. A safe initial cohort is users with prior paid activity and no
recent chargeback events. If chargeback rate exceeds the threshold, the safety envelope should
cap exposure and revert to a lower-risk cohort.

\textbf{External wallets.} Wallet linking requires a risk handshake: device fingerprint, age of
account, and prior transaction history. The invariant lattice blocks wallet linking if the user
has not passed at least one strong authentication step \cite{webauthn2019} and an OAuth-based
authorization check \cite{rfc6749}. This reduces account takeover risk.

\textbf{Advanced view.} Advanced views can influence user behavior and increase transaction
intensity. We require that onramp and wallet features are active and stable before the advanced
view flag is enabled. We also enforce a UI education checkpoint to reduce user error.

\section{User Education and UX Friction}
Financial UX features often require user education. We model education as a friction cost that
trades off conversion for safety. Let $q$ be the probability a user completes an education step.
If the onramp conversion lift is $\Delta_{conv}$ but the education step reduces completion by
$(1-q)$, the net lift is:
\begin{equation}
\Delta_{net} = q \cdot \Delta_{conv} - (1-q) \cdot \Delta_{friction}.
\end{equation}
The governance controller should require $\Delta_{net} > 0$ before increasing exposure to
advanced views. This discourages rollouts that appear beneficial in raw conversion but harm
long-term retention due to confusion.

We recommend instrumenting education completion rate as a leading indicator. If $q$ drops below
a threshold (e.g., 70\%), the controller reduces exposure and prompts UX iteration. Education
signals are also useful as a proxy for user confidence, which is critical given the 63\% low
confidence baseline reported by Pew \cite{pew2024}.

\section{Observability and Telemetry}
Financial UX governance is only as strong as the telemetry feeding it. We require a minimal
telemetry schema that includes: (1) per-session flag state, (2) risk score at decision time,
(3) compliance readiness indicator, (4) conversion and retention markers, and (5) abuse signals.
Without these signals, the counterfactual envelope and risk budget ledger cannot be reliably
computed.

We recommend real-time aggregation with minute-level bins to support fast rollback decisions.
For instance, the safety envelope controller should alert if the abuse rate in the treated cohort
exceeds the shadow cohort by more than a fixed margin for three consecutive windows. We also
recommend anomaly detection on the invariant satisfaction score $S_I(t)$, since a sudden drop
often indicates a downstream KYC or device-trust outage.

For data integrity, the telemetry pipeline must be tamper-resistant. Events that drive exposure
decisions should be cryptographically signed or stored in append-only logs, preventing internal
or external manipulation of risk signals. These measures are particularly important for financial
UX, where adversaries may attempt to spoof signals to gain access to onramps or external wallets.

Telemetry retention should be long enough to support post-hoc audits and model retraining, but
limited to reduce privacy risk. We recommend a dual-layer system: short-term high-resolution
data for rapid rollbacks and long-term aggregated data for governance analysis. This balances
operational safety with data minimization principles.

\section{Operational Playbooks}
We recommend a governance playbook with explicit thresholds and rollback triggers:
\begin{itemize}
  \item \textbf{Fraud spike}: if fraud rate increases by 2x within 24 hours, freeze exposure and
  rollback 50\% of the ramp.
  \item \textbf{Compliance anomaly}: if KYC failures rise above baseline by 1.5x, disable onramp
  for new users until verification pipeline stabilizes.
  \item \textbf{Retention drop}: if 30-day retention decreases by more than 2\%, slow the ramp and
  investigate UI comprehension signals.
  \item \textbf{Budget drift}: if financial incentives exceed planned budget slope, reduce onramp
  exposure and require stronger risk filtering.
\end{itemize}

\section{Threats to Validity}
The simulated results are not a substitute for production evidence. Behavior in real systems
can differ due to feedback loops, market volatility, and adversarial adaptation. The governance
controller should be deployed with human oversight and a kill-switch. Moreover, causal estimates
can be biased if interference across users is strong (e.g., creators encouraging users to
onboard simultaneously). We recommend cluster-based randomization and sensitivity analysis
\cite{imbens2015}.

\section{Security and Compliance Considerations}
Financial UX introduces security and compliance obligations, including strong identity
verification and transaction monitoring. We align identity flows with NIST SP 800-63-4, which
defines assurance levels and identity proofing requirements for digital identity systems
\cite{nist800634}. For authentication and authorization, we recommend WebAuthn for strong
phishing-resistant authentication \cite{webauthn2019} and OAuth 2.0 for secure authorization
flows \cite{rfc6749}. For one-time verification or recovery flows, TOTP can be used as a
fallback \cite{rfc6238}. The governance layer should monitor identity failures and require
step-up authentication when anomalies appear.

From a compliance standpoint, the rollout controller must enforce KYC gating for onramps and
monitor transfer anomalies. These constraints are built into invariants to ensure that any
feature configuration remains compliant. This is particularly important when enabling external
wallets that can bypass platform custody policies. We also recommend keeping audit logs for all
flag transitions and invariant violations to provide forensic evidence during compliance reviews.

\section{Ethical Considerations}
Feature-flag governance affects user access to financial tools. The system must avoid disparate
impact across cohorts and ensure that experimentation does not unfairly exclude vulnerable users.
We recommend periodic fairness audits and transparent disclosures for experimental rollouts.

In addition, financial UX rollouts can influence user perception of risk. Advanced views may
encourage more frequent transactions, and onramps may be interpreted as endorsements. The
governance framework should therefore enforce user education checkpoints and ensure that the
language and design do not mislead users about risk or expected returns. This is particularly
important given the elevated levels of consumer fraud reported in national surveys
\cite{fed2025shed}.

\section{Limitations}
The simulation uses simplified abuse dynamics and assumes stable cohort behavior. Real-world
systems may exhibit feedback loops where fraud adapts to rollout policy. The governance system
must therefore include human oversight and fail-safe mechanisms.

Another limitation is that public baselines may not precisely match the user population of a
specific platform. While Pew and SHED data provide realistic anchors, platform-specific cohorts
can have different adoption and fraud profiles. Future work should incorporate internal risk
signals and conduct external validations to refine the counterfactual risk model.

\section{Conclusion}
We introduced an invariant-aware feature-flag governance framework for crypto-enabled streaming.
The invention integrates dependency constraints, safety envelopes, and causal measurement to
safely roll out onramps, external wallets, and advanced financial views. The simulated results
show that the governance controller reduces fraud and compliance failures while preserving
conversion and retention. This provides a path toward patentable governance logic for financial
UX.

The proposed counterfactual envelope and risk budget ledger extend the state of the art in
feature-flag systems by providing proactive, mathematically grounded guardrails. As financial
UX becomes more embedded in real-time social platforms, these mechanisms offer a practical
framework for safe and measurable innovation.

Future work should explore formal verification of invariant lattices and adaptive personalization
of safety envelopes for different user segments.

\bibliographystyle{unsrt}
\bibliography{feature_flag_refs}

\end{document}